\documentclass[prx,twocolumn,floatfix]{revtex4}
\usepackage[colorlinks=true,citecolor=blue,linkcolor=blue,urlcolor=blue]{hyperref}
\usepackage[sort&compress]{natbib}
\usepackage{graphicx,times}
\usepackage{color} 
\usepackage{sansmath}
\usepackage{amssymb}

\usepackage{subeqnarray}
\usepackage{diagbox}
\usepackage{makecell}
\usepackage{amsmath}
\usepackage{mathtools}
\usepackage{empheq}

\usepackage{enumitem}

\usepackage[dvipsnames]{xcolor}

\definecolor{aprcolor}{rgb}{0.7,0.1,0.2}

\newcommand{\mathbbm}[1]{\text{\usefont{U}{bbm}{m}{n}#1}}

\DeclareSymbolFont{greekletters}{OML}{FiraSans}{m}{n}

\begin{document}
\title{On ``Consistent Quantization of Nearly Singular Superconducting Circuits"}
\author{I. L. Egusquiza}
\email{inigo.egusquiza@ehu.es}
\affiliation{Department of Physics, University of the Basque Country UPV/EHU, Apartado 644, 48080 Bilbao, Spain}
\affiliation{EHU Quantum Centre, University of the Basque Country UPV/EHU, Apartado 644, 48080 Bilbao, Spain}
\author{A. Parra-Rodriguez}
\email{ adrian.parra.rodriguez@gmail.com}
\affiliation{Instituto de Física Fundamental IFF-CSIC, Calle Serrano 113b, 28006 Madrid, Spain}

\begin{abstract}
The analysis conducted by Rymarz and DiVincenzo~\cite{Rymarz:2023} regarding quantization of superconducting circuits is insufficient to justify their general conclusions, most importantly the need to discard Kirchhoff's laws to effect variable reductions. Amongst a variety of reasons, one source of several disagreements with experimental and theoretical results is  the long-standing  dispute 
 between extended vs compact variables in the presence of Josephson junctions.

\end{abstract}

\maketitle
Recently, Rymarz and DiVincenzo in~\cite{Rymarz:2023} (henceforward referred to as RD) have questioned the correctness of standard canonical quantization methods based on variable reduction using Kirchhoff's laws. {They} based {their conclusions} on {a perceived} mismatch of a {classical description of a singular {electrical} circuit, on the one hand,} and a quantum adiabatic approximation {arising from a regularization of the circuit, on the other}. {This regularization, frequently taken for granted in classical circuit analysis, consists on considering}  parasitic capacitances {that are then} taken to zero. The question raised by RD is extremely interesting, relevant and timely, given that it sets the grounds for the more general question of how to obtain reduced quantum Hamiltonian descriptions of nonlinear superconducting circuits. RD explicitly reads ``we conclude that classical network synthesis techniques which build on  Kirchhoff’s laws must be critically examined prior to applying circuit quantization", a conclusion that is restated in several places, e.g. ``Our results
justify the conclusion that one should not use Kirchhoff’s
current law to eliminate variables in the Lagrangian."  Here we show that this conclusion does not follow from their computation. Furthermore, it contradicts previous analyses that match experimental results. Additionally, there are implicit assumptions in RD that have otherwise been challenged from both  a theoretical and a phenomenological perspective.

This conclusion in RD is not restricted to nonlinear circuits. Thus, in their final Section V, they  dispute linear circuit theory reduction as applied to superconducting circuits: ``To conclude, our work highlights the importance of critically examining the validity of well-tried theorems or simplifications from classical network synthesis, which build on Kirchhoff’s conservation laws (e.g., the Y-$\Delta$ transformation or Tellegen’s replacement rules for terminated gyrators [58,96]), prior to applying circuit quantization" (references as in the original, here \cite{RymarzPhD:2023} and \cite{Tellegen:1948}).  Notice that this statement partially contradicts  a previous one: ``In particular, for a quantized description of the system,
such an elimination of variables on the Lagrangian level should be handled with care, and, strictly speaking, is valid for linear systems only. The classical elimination of variables, motivated by the singularity of the Lagrangian, must be deemed an incorrect procedure as it completely misses quantum-fluctuation effects, which become large as the singular limit is approached".

Here, we challenge these and several other unsupported statements made in Ref.~\cite{Rymarz:2023}. Some are already beginning to propagate through the literature, see for instance citations of RD in~\cite{Miano:2023,Ciani:2024,Willsch:2024}. To boot, in other works (not referenced) there is a clear  misunderstanding of the conclusions in RD, which are wrongly taken to be a solution for  quantization in general in the presence of singularities.

\section{Context and brief review of Ref.~\cite{Rymarz:2023}}

\begin{figure}[!ht]
	\includegraphics[width=1
\linewidth]{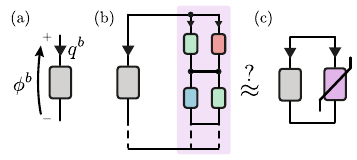}
	\caption{(a) In electrical circuit theory, a generic two-terminal (1-port) lumped element is described by two variables: a branch charge and a branch flux, and a constitutive (possibly nonlinear and nonlocal in time) relation between them \cite{Chua:1980}. (b) When working with many-element circuits, there is a strong general interest in obtaining approximating low-energy descriptions with (c) fewer elements. For instance, in superconducting circuits, low-energy models with few nonlinear lumped elements have been used~\cite{Manucharyan:2009,Yan:2020,Ye:2021} when capacitive parasitic effects have been neglected altogether~\cite{Miano:2023}. 
 }
	\label{fig:Reduced_LE_models}
\end{figure}

Lately, there has been an enhanced interest in developing reduced ($N-k$)-body models of initially $N$-body descriptions of lumped-element nonlinear superconducting circuits~\cite{Kitaev:2006,Dempster:2014,Viola:2015,Yan:2020,Smith:2020,Miano:2023}, see Fig.~\ref{fig:Reduced_LE_models}. The reduction problem was one of the driving forces in the development of analytical mechanics and pervades all applications of quantum mechanics as well. Yet the specificity of superconducting circuits presents its own challenges, and its full understanding would lead to huge advances in the field. A   particular subclass of problems of current interest in this regard involves the smooth removal of a subset of linear electric elements of the full circuit.  Specifically, parasitic capacitances and inductances. The removal is understood as either infinite or zero limits, depending on whether they are in parallel or series connections. Standard examples of such procedures can be found in low-energy descriptions of the \emph{fluxonium}~\cite{Manucharyan:2009}, the \emph{quarton}~\cite{Yan:2020,Ye:2021}, the \emph{KITE}~\cite{Smith:2020}, etc. 

Focusing on the parasitic capacitance problem and using the quantum Born-Oppenheimer approximation, in Section III RD asserts having determined generic 1-body low-energy effective quantum Hamiltonians for a family of 2-body regularized nonlinear circuits that contradict the quantization of classical singular circuits. From this  claim of  disagreement they move to challenge the application of Kirchhoff's constraints to reduce the number of degrees of freedom for electrical circuits, if these are to be working in the quantum regime. 

\begin{figure}[h]
	\includegraphics[width=1
\linewidth]{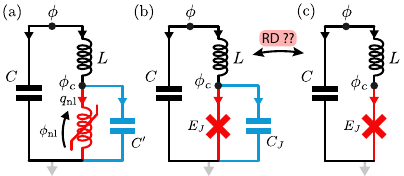}
	\caption{(a) The main results of RD 
  (Theorems 1-4) pertain to the limit $C'\rightarrow 0$ for different classes of nonlinear inductors, all of them described by nonlinear constitutive relations between their branch charge $q_{\text{nl}}$ and flux $\phi_{\text{nl}}$. (b) According to the classification in RD, the pure Josephson element (with its parallel parasitic capacitance $C_J$), described by the nonlinear relation $\dot{q}_J=E_J\left(\frac{2\pi}{\Phi_Q}\right)\sin\left(\frac{2\pi \phi_J}{\Phi_Q}\right)$, is a type 1a inductor. 
  Consequently, Theorem 1 should be applicable. RD explicitly shows that the circuit in (c), i.e., (b) without the Josephson parasitic capacitance, is  singular in two ways. First, in having a defective kinetic term in its flux presentation. Second, in that  obstructions to variable reduction exist for some parameter regimes of $L$ and $E_J$. They attribute this failure to the Dirac-Bergmann algorithm.}
	\label{fig:RD_Circuit}
\end{figure}

The regularized class of circuits they study  consists of a series connection of a capacitor ($C$), an inductor ($L$), and a nonlinear inductor (described by a nonlinear charge-flux relation $\dot{q}_{\text{nl}}=\partial U_{\text{NL}}(\phi_{\text{nl}})$) with an attached parasitic parallel capacitance $C'$, whose limit to zero is to be researched, see Fig.~\ref{fig:RD_Circuit}(a).

Indeed, they motivate such a general scenario in a previous section (II), from the quantization analysis of  a a family of circuits.  In this specific problem, the nonlinear inductor is a Josephson junction ($E_J$), which is typically modelled with a parasitic parallel capacitance \footnote{A more sophisticated model of the Josephson junction considers a parallel dissipative element in parallel with the other two elements~\cite{McCumber:1968}, but it is not part of the discussion of Ref.~\cite{Rymarz:2023}.}, usually denoted as $C_J \equiv C'$, see Fig.~\ref{fig:RD_Circuit}(b). When $C_J=C'=0$, and  for some parameter regimes, this family includes    circuits that are singular also from a classical analysis perspective~\cite{Chua:1980}, see Fig.~\ref{fig:RD_Circuit}(c). This purely mathematical question is very well motivated by the fact that such a capacitance can potentially be designed to be orders of magnitude smaller than the other one ($C_J/C \ll 1$), for instance, in the \emph{transmon}~\cite{Koch:2007}. RD  finds the process wanting, as this is in fact a long-known unsolvable singular problem in quantum circuits for some parameter range, see for instance Refs.~\cite{Ilichev:2001,Golubov:2004, Kafri:2017}. From a classical circuit modelling perspective, the circuit in Fig.~\ref{fig:RD_Circuit}(c) is  ill-posed~\cite{Chua:1980} when the screening parameter 
\begin{align}\label{eq:betadef}
    \beta\equiv L E_J\left(\frac{ 2\pi}{\Phi_Q}\right)^2>1,
\end{align}
with $\Phi_Q=h/2e$  the superconducting magnetic flux quantum.

In the face of this issue, RD invokes the Dirac-Bergmann (DB) algorithm as an additional technique to the standard node-flux quantization method, and claim ``its failure" \footnote{Naturally, the alternative, and equivalent method of Faddeev-Jackiw~\cite{ParraRodriguez:2023} ``fails" as well, as they properly mention.}. In brief, the problem with this circuit is that, if one eliminates the parasitic capacitance  completely and studies the equations of motion, there exists a consistency condition associated to the node where the Josephson junction (JJ) and the inductor meet, namely  Kepler's equation
\begin{align}\label{eq:betaeq}
    \varphi-\varphi_c = \beta \sin \varphi_c,
\end{align}
which, for $\beta>1$, imposes an uninvertible relation  between the node phases $\varphi_c$ and $\varphi$. Here, $\varphi_\alpha=2\pi\phi_\alpha/\Phi_Q$ are the phase variables, proportional to the flux variables. The physical interpretation of this consistency condition is that it follows from   current conservation  at that node. To be clear, for its assertion one needs both the constitutive relations and Kirchhoff's current law. In the case at hand, this consistency condition can be identified directly from the equations of motion and does not require the DB method nor any equivalent (Faddeev-Jackiw~\cite{Faddeev:1988}) one. Nonetheless,  at this point RD concludes that there is a  ``failure of the Dirac-Bergmann algorithm when applied to
electrical networks". It also asserts  that ``[o]ur results justify the conclusion that one should not use Kirchhoff’s current law to eliminate variables in the Lagrangian". 

To justify this conclusion, in Section III RD studies \emph{quantum mechanical} reduction of a regularized  Hamiltonian  for all parameter regimes, and for  even more general nonlinear inductors, see Fig.~\ref{fig:RD_Circuit}(a) \footnote{Observe that, as they mention, their  set of nonlinear inductors studied is not complete: ``Note that not all possible nonlinear inductors can be classified into one of the three types we provide".}. That is, they pose the question whether the parasitic capacitance limit $C'/C\rightarrow 0$ could be smoothly taken quantum mechanically. 

 Rymarz and DiVincenzo approach the question in the context of the Born-Oppenheimer (BO) approximation. In doing so, they claim to have identified generic behaviours, with corresponding  fixed points in a renormalization picture. To be  precise, they study the following four types of inductors:
\begin{itemize}
	\item type~1 (sublinear):
	  \begin{enumerate}
		\item[] (a) $U_{\text{NL}}(\phi_{\text{nl}}) = U_{\text{NL}}(-\phi_{\text{nl}})$ 

        $\exists\,\gamma\in(0,2): \lim_{\phi_{\text{nl}} \to \pm \infty} U_{\text{NL}}(\phi_{\text{nl}})/\phi_{\text{nl}}^{\gamma} = 0 $
		\item[] (b) $U_{\text{NL}}(\phi_{\text{nl}}) \neq U_{\text{NL}}(-\phi_{\text{nl}})$ 

        $\exists\,\gamma\in(0,1): \lim_{\phi_{\text{nl}} \to \pm \infty} U_{\text{NL}}(\phi_{\text{nl}})/\phi_{\text{nl}}^{\gamma} = 0 $
	\end{enumerate}
	
	\item type~2 (superlinear): \\
	$\lim_{\phi_{\text{nl}} \to \pm \infty} \phi_{\text{nl}}^2/U_{\text{NL}}(\phi_{\text{nl}}) = 0$
	
	\item type~L (quasilinear): \\
	$\exists\,\mathfrak{L}>0: \lim_{\phi_{\text{nl}} \to \pm \infty} U_{\text{NL}}(\phi_{\text{nl}})/\phi_{\text{nl}}^2=1/{2 \mathfrak{L}}$ 
	\quad and  \\
	$U_{\text{NL}}(\phi_{\text{nl}}) - \phi_{\text{nl}}^2/2\mathfrak{L}$ describes a type-1 inductor
\end{itemize}
We have rewritten their list in Sec. III (b) using branch variables, which are the defining variables of lumped elements~\cite{Chua:1980}, instead of the node flux variables as in RD. We do so to emphasize that the  node flux (and loop charge) variables are introduced to solve those very Kirchhoff laws that RD challenges.

Indeed, it is only through the requirement of Kirchhoff constraints that RD arrives at Lagrangian (14) and Hamiltonian (15) (referenced as in RD), which we rewrite here
\begin{align}
    L&=\frac{C\dot{\phi}^2}{2} + \frac{C'\dot{\phi}_c^2}{2} - \frac{(\phi-\phi_c)^2}{2L} - U_{\text{NL}}(\phi_c),\label{eq:L_RD_circuit}\\
    H_r&=\frac{Q^2}{2C} + \frac{Q_c^2}{2C'} + \frac{(\phi-\phi_c)^2}{2L} + U_{\text{NL}}(\phi_c),\label{eq:Hr_RD_circuit}
\end{align}
where $Q$ and $Q_c$ are the conjugated charges to the node fluxes $\phi$ and $\phi_c$.

In application of the first elements of the BO study, RD sets out a \emph{fast} and a \emph{slow} Hamiltonian as parts of Eq. \eqref{eq:Hr_RD_circuit}. In the choice made in RD (see below for details), the fast Hamiltonian still depends on the small parameter $\kappa^4\equiv 
C'/C$ (our notation). The crucial analysis in RD, technically brilliant, is focused on the estimation of the ground state energy of the fast Hamiltonian and its dependence on $\phi$. The main hurdle is the possible nonanalyticity of the fast Hamiltonian in the parameter $\kappa$. Under some explicit and some \emph{implicit} assumptions, 
RD states that Theorems 1-4 provide adiabatic low-energy Hamiltonian approximations of (\ref{eq:Hr_RD_circuit}) when   $\kappa^4=C'/C\rightarrow 0$, namely 
\begin{align}
    H_{r, \text{eff}}=\frac{Q^2}{2C}+U_{\text{BO}}(\phi).\label{eq:Th1_Hr_eff}
\end{align}

We will qualify that statement below.

Specifically,  the content of Theorem 1 is that for type 1(a) potentials the  smallest eigenvalue of the fast Hamiltonian does not depend on $\phi$ in the limit $\kappa\to0$, and Theorem 2 extends that conclusion to type 1(b) potentials. We will \emph{not} dispute the theorems, but rather be explicit about assumptions. Hence they conclude that $U_{\mathrm{BO}}(\phi)$, defined through that eigenvalue, is zero.

\begin{figure}[!ht]
	\includegraphics[width=1
\linewidth]{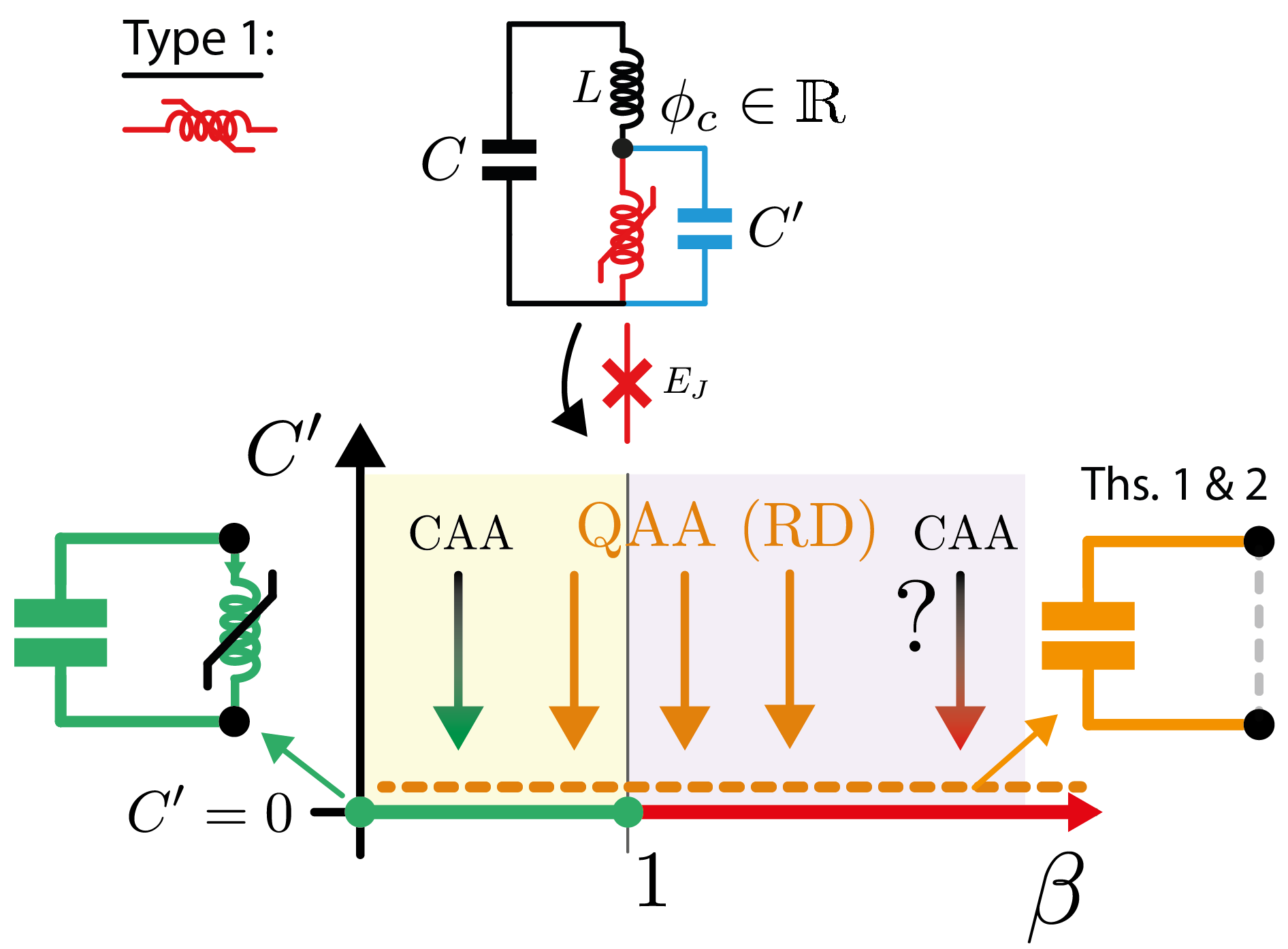}
	\caption{RD performs a quantum adiabatic approximation (QAA) to the  circuit in the top of the figure by taking the limit \(C' \rightarrow 0\) while assuming that the phase space before quantization is \(T^*\mathcal{X} = \mathbb{R}^4\), which includes \(\phi_c \in \mathbb{R}\). RD asserts that the application of their Theorems 1 and 2 (Theorem 1 includes the unbiased Josephson junction, while Theorem 2 involves non-symmetric bounded potentials, such as SQUIDs and SNAILs) provides a general effective low-energy Hamiltonian dynamics approximated by the orange circuit for inductors of type 1, independent of the other parameters ($C$ and $L$). They also assert that such a limit is in strong contradiction to the specific case of a Josephson junction when the capacitance has been removed before the quantization (\(C' = 0\)), resulting in the effective circuit model in green for \(\beta \equiv L E_J (2\pi/\Phi_Q)^2 \leq 1\). We  show in the main text that for \(\beta < 1\): (i) there is no such contradiction between the two models under their assumption, and (ii) a simple classical adiabatic approximation (CAA, black-to-green arrow) also renders the same model.
}
	\label{fig:Beta_sketch}
\end{figure}

In other words, RD claims that, for the {single} JJ case {(which falls in the type 1 class)}, and independently of the value of $\beta$, the low-energy spectrum when $\kappa^4= C'/C\rightarrow 0$ is well approximated by the Hamiltonian $H_{r,\text{eff}}=\frac{Q^2}{2C}$, with a continuous spectrum. In circuital representation, they claim their approximated model to be the orange circuit in Fig.~\ref{fig:Beta_sketch}. 

{At a first glance,} this could indeed appear to be a striking contradicting statement if one naively compares it with RD's results in Sec. II (and others, e.g., ~\cite{Ilichev:2001,Golubov:2004,Richer:2017,Smith:2023,Willsch:2024}), where the application of variable reduction for the case of $\beta\leq 1$ leads to a different, yet perfectly quantizable, nonlinear Hamiltonian model (see their Eq.~(12), equivalent to the green circuit in Fig.~\ref{fig:Beta_sketch})
\begin{align}
    H_s=\frac{Q^2}{2C}+\frac{[\phi-\phi_c(\phi)]^2}{2L}+U_{\text{NL}}[\phi_c(\phi)].\label{eq:H_s_RD}
\end{align}

In their own words: ``In the single-valued case ($\beta\leq 1$), the Hamiltonian $H_s$ in
Eq.~(12) is a mathematically well-defined function of a pair
of two conjugate variables, and it can be used in the usual
way to describe the dynamics of the system [57,61]. In particular, a quantized description is obtained by promoting
the canonical variables to operators and imposing the
canonical commutation relation $[\phi,Q]=i\hbar$." According to RD, the reason between this supposed mismatch is due to quantum fluctuations: ``The classical elimination of variables, motivated by the singularity of the Lagrangian, must
be deemed an incorrect procedure as it completely misses quantum-fluctuation effects, which become large as the
singular limit is approached."

Based on the  findings we  summarized above, they go on to question the whole framework in which these models have been developed, namely circuit theory, and in particular, the way in which variable reduction is typically applied in superconducting circuits in the quantum regime in the presence of nonlinear (Josephson junction) elements. 

In this text, we demonstrate that:

\begin{itemize}
    \item The singular behaviour for $\beta>1$ in the consistency equation \eqref{eq:betaeq} is recovered from a classical adiabatic limit of the regularised classical circuit. What RD terms a failure of DB is a reflection of the underlying classical critical behaviour (existence of a bifurcation controlled by the $\beta$ parameter).
    \item There is an implicit assumption in the theorems of RD, namely that $\phi_c$ extends to the full real line (extended variable in what follows). The issue of whether flux variables in the context of superconducting circuits can be compact or are always extended has been, and still is hotly debated. We show that when $\phi_c$ is a compact variable, the classification of potentials in RD is inapplicable,  and the  theorems do not follow nor do the conclusions drawn from them in RD.
    \item Under the assumption of extended variable, for the case of a cosine potential we show that the quantization of the classical adiabatic limit, when computable,  is in agreement with Theorem 1 of RD. This theorem only provides us with spectral information, and an identically zero potential is not the only one with a low-energy continuous spectrum. 
    \item Thus the theorems in RD do not contradict the applicability of Kirchhoff's laws for reduction and then quantization even for the restricted context in which they are applicable. The conclusions in RD do not follow from the evidence presented there, not for the concrete family under analysis nor much less in fuller generality.
    \item Experimental evidence is not aligned with the conclusions in RD for a variety of systems.
\end{itemize}

We also note, although we do not investigate fully the consequences, that the theorems in RD only pertain to the ground state of what has been termed the fast Hamiltonian, and no analysis of nonadiabatic transition elements has been carried out.

To present these points more fully, we shall begin with a summary of the current discussion on the compact or extended character of flux variables in superconducting circuits, that we present in Section \ref{sec:atale}. We then review the context of the theorems for extended variables, Section  \ref{sec:RDs_Th1_R}, making special emphasis on what changes of scale really are. We then compute the classical adiabatic limit using the slow manifold approach, and derive the effective consistency condition to lowest order in the small parameter $\kappa$. We demonstrate that it is the consistency condition that must be met were we to start in the directly reduced model, i.e., with $\kappa$ being directly set to zero. For the regular consistency condition we give the explicit parametric presentation of the classical effective potential. At this point we draw a first set of conclusions. First, DB (or equivalently Faddeev-Jackiw) is in these cases detecting the failure of the \emph{classical} adiabatic limit. Second, for the case of a cosine potential (amongst others) the quantization of the classical effective Hamiltonian gives rise to a low-energy  continuous spectrum, consistent with theorem 1. We then pass on the subtler issue of compact variables in Section \ref{sec:No_Th1_S1}, and provide several arguments regarding the inapplicability of the central theorems of RD to this situation. As we have detected some confusion in the literature regarding changes of scale in that context we give a full account of the nonexistence of canonical dilation transformations in that case, both classically and quantum mechanically.  Next, in Sec. \ref{sec:Adiabatic_S1}, we compute the classical adiabatic limit for compact $\phi_c$, and present the phenomenon of classical effective compactification. The comparison to the quantum case is only sketched, as there are in fact open problems (not covered in RD) in this regard. As the ultimate goal of this endeavour is the construction of good quantum models for actual devices, in Sec. \ref{sec:Experimental_evidence} we discuss relevant experimental evidence and phenomenological approaches to it. We conclude in a Summary in Sec. \ref{sec:Summary}.

\section{One tale, two viewpoints, and many narrators}\label{sec:atale}
There has been a long-standing and ongoing debate regarding the spectrum of the macroscopic flux and charge operators in superconducting circuits~\cite{Rogovin:1982,Widom:1982,Averin:1985,Likharev:1985,Zwerger:1986,Buettiker:1987,Apenko:1989,Zaikin:1990,Schoen:1990,Mullen:1993,Bouchiat:1998,Koch:2009,Averin:2018,ThanhLe:2020,Murani:2020,Hakonen:2021,Murani:2021,Devoret:2021,Sonin:2022}, or equivalently, the topology of their classical counterparts. This debate is central to understanding the matching of classical dynamics (classical limit) described by Kirchhoff's equations with quantum mechanics.

There have been two main viewpoints on the matter. On the one hand, a historical presumption held that all macroscopic (charge and flux) quantum observables (thus operators) have a  spectrum that covers the full real line, independent of the presence or absence of superconducting islands. This view can be found, for example, as early as in Ref.~\cite{Widom:1982}. On the other hand, some have suggested that certain of these operators should have an integer spectrum, and consequently, their conjugate counterparts a compact spectrum. This perspective was also discussed around the same time, see Ref.~\cite{Rogovin:1982}. This latter viewpoint has been prominent since the first theoretical explanations for quantum-coherent experiments involving island-based superconducting qubits towards the end of the past century. For instance, in the Cooper-pair box~\cite{Nakamura:1999}, the Cooper-pair number operator was posited to have an integer spectrum by Bouchiat et al.~\cite{Bouchiat:1998}, and therefore, its conjugate phase variable was considered to have a compact character.

Several different phenomenological and mathematical explanations to assume one or the other have been put forward, and  even non-binary perspectives have been used~\cite{Murani:2020}, intensely reigniting the debate~\cite{Hakonen:2021, Murani:2021, Sonin:2022}. Interestingly, there have been theoretical attempts to reconcile these perspectives~\cite{Apenko:1989,Schoen:1990,Koch:2009}, as well as experimental proposals  (for now, purely  \emph{gedanken} experiments) to tell them apart~\cite{ThanhLe:2020}. 

In what concerns us here, the analysis of RD ignores this discussion in their Secs. I, II, and III, where the connection is drawn between the supposed failure of DB for the singular circuit in Fig.~\ref{fig:RD_Circuit}(c), and the theorems in RD as applied to for Fig.~\ref{fig:RD_Circuit} (a).  Of course, this is the central argument of RD to discard Kirchhoff's equations for variable reduction. Indeed, RD analyzes the family of circuits in their Fig. 4 of Sec. III assuming classical phase space to be $\mathbbm{R}^4$, and that under canonical quantization, therefore,  all the quantum operators have real spectrum, including the node-flux $\phi_c$. 

This assumption has been applied, also implicitly, to the Josephson case in RD. In RD Sec. I we read : ``The Josephson case is in the sublinear universality class and flows to the open-circuit fixed point". As indicated above, we understand this statement in the context of the circuit of RD Fig. 4, given that the classification pertains to the circuital node-flux variable $\phi_c$, rather than to the branch flux variable for the nonlinear element. Moreover, the discussion on the compact vs extended flux variables is also missing in Sec. II for the ``singular" circuit, see Fig.~\ref{fig:RD_Circuit}(c) (Fig.~3 in RD), although it is implicitly assumed that the reduced phase-space for $\beta\leq 1$ is $\mathbbm{R}^2$. Here it should be noted that this circuit indeed has two disconnected superconducting islands.

Interestingly, and surprisingly, RD does take a different point of view in Sec. IV, where  a different circuit  is analyzed, consisting of a loop of three Josephson junctions, i.e., three disconnected superconducting islands. There, they recall the oft-repeated argument that the symmetries of the Hamiltonian function (in a particular flux-charge basis) determine/allow both extended or compact descriptions (e.g., see Ref.~\cite{Murani:2020}): ``The Josephson junction has an additional feature that we do not treat above, namely, that its potential characteristic is periodic. This has the consequence that the flux variable $\varphi$ can be treated as compact on the domain $(0, \Phi_0]$, leading also to the charge on the nodes of the junction being constrained by the uncertainty relationship $\Delta\phi \Delta Q \approx \hbar$."

In summary, making general statements about the validity of classical reduction previous to quantization in the context of superconducting circuits without assessing their applicability and impact on compact variables is at least premature in the current stage of the scientific discussion.

\section{On Theorem 1 (and 2) with $\phi_c\in \mathbbm{R}$}
\label{sec:RDs_Th1_R}
In this section, we will review the analysis of RD for the circuit in Fig.~\ref{fig:RD_Circuit}(a) for the class of type 1a inductors (related to Theorem 1), quantum mechanically, and compare it to a classical adiabatic analysis for $\beta<1$. Our conclusions naturally extend to the other sublinear inductor case studied in RD, namely, type 1b, but it will be  omitted for the sake of simplicity. 

\subsection{Quick review of the proof}
Let us start with the relevant family of Hamiltonians in RD, Eq.~(\ref{eq:Hr_RD_circuit})  (Eq.~(15) of \cite{Rymarz:2023}). In order to facilitate the discussion, RD introduced new symbols for the $LC'$-resonator frequency and the corresponding flux zero-point fluctuation,
\begin{align}
    \omega'_r=\frac{1}{\sqrt{LC'}}\,,\qquad \Phi_{\mathrm{ZPF}}=\left(\frac{\hbar^2L}{C'}\right)^{1/4}\,.
\end{align}
These are used to adimensionalize variables $\phi_c=\Phi_{\mathrm{ZPF}} y$ and $Q_c= \hbar p_y/\Phi_{\mathrm{ZPF}}$, and the symplectic two-form has the local expression into 
\begin{align}
    \mathrm{d}Q_c\wedge\mathrm{d}\phi_c=\hbar\mathrm{d}p_y\wedge\mathrm{d}y\,.
\end{align}
Notice that using the symplectic form in this case is equivalent to the Poisson bracket, and we simply use this form to better keep track of the dimensional aspect when required.

RD further introduces the \emph{dimensionful} variable $\epsilon=1/\sqrt{\hbar\omega'_r}$ which they take as a small variable. It is however convenient, in setting up a Born--Oppenheimer analysis, to identify clearly the small \emph{adimensional} parameter that controls the approximation. In the case at hand, we introduce the relevant small adimensional parameter as
\begin{align}
    \kappa=\left(\frac{C'}{C}\right)^{1/4}\,.
\end{align}
The choice of exponent follows  that in the BO expansion for molecules, in connection with analyticity properties in $\kappa$ of all magnitudes. Even though here we do not carry out a detailed analysis, we will see below that it is a convenient choice, in particular in  expression \eqref{eq:adimHr}.

This adimensionalization step seems uncontroversial at this moment, but we shall later see that there are subtleties involved. Let us forge ahead, and, for clarity, we   adimensionalize the variables $\phi$ and $Q$ as well. An  important point to bear in mind is that the adimensionalization of these variables need not be carried out with the same dimensionful constants as before, namely $\Phi_{\mathrm{ZPF}}$ for fluxes and $\hbar/\Phi_{\mathrm{ZPF}}$  for charges at the moment. 

Thus let us  also define the zero-point fluctuation flux magnitude for the $LC$-oscillator and its corresponding frequency, 
\begin{align}
   \Phi_C= \left(\frac{\hbar^2L}{C}\right)^{1/4}, \,\qquad \omega_C= \frac{1}{\sqrt{LC}}\,,
\end{align}
as well as a dimensionful parameter
\begin{align}
   \epsilon_C=\frac{1}{\sqrt{\hbar\omega_C}} \,,
\end{align}  
and introduce adimensional variables $x$ and $p_x$ defined by 
\begin{align}
    \phi= \Phi_C x\,,\qquad Q= \frac{\hbar}{\Phi_C} p_x\,,
\end{align}
such that the symplectic two-form is 
\begin{align}
    \mathrm{d}Q\wedge\mathrm{d}\phi=\hbar\,\mathrm{d}p_x\wedge\mathrm{d}x\,.
\end{align}

We then rewrite the Hamiltonian in the new set of adimensional variables as
\begin{align}
\label{eq:adimHr}
    H_r& = \frac{1}{\epsilon^2}\left[\frac{\kappa^2}{2} p_x^2 +\right.\nonumber\\
    &  \left. \frac{1}{2}p_y^2+\frac{1}{2}\left(y-\kappa x\right)^2 + \kappa^2 \epsilon_C^2 U_{\mathrm{NL}}\left(\frac{y\Phi_C}{\kappa}\right) \right]\,.
\end{align}
It is this form that we shall now investigate. We stress that we have not deviated from RD to this point in anything but in the introduction of some new dimensionful constants and one small adimensional parameter $\kappa$. The reason to do so is that we now clearly identify that the limit that we study is $\kappa$ to 0 in Eq. \eqref{eq:adimHr}, with all other parameters inside the square brackets constant. The astute reader will have noticed that the global energy scale, $1/\epsilon^2$, does depend on $C'$ (unlike the parameters inside the square brackets, other than $\kappa$). Yet it is the \emph{global} energy scale, in that it affects equally all terms of the Hamiltonian, and the BO approximation concerns \emph{separation of scales} in the Hamiltonian.

We can now attempt some initial classical intuition about the limit $\kappa\to0$. Namely, if $U_{\mathrm{NL}}$ is sublinear according to the definitions of RD (Subsection III.B), in the limit $\kappa\to0$ it is intuitive that it would have no effect on the dynamics. For example, consider $U_{\mathrm{NL}}$ sinusoidal. Then as $\kappa\to0$ it seems to disappear. This rough qualitative analysis is, as we shall see, too na\"{\i}ve to allow us to stop here.

In particular, we can identify the \emph{slow} kinetic term $\kappa^2p_x^2/2$, and one can choose to take the rest of the (adimensional) hamiltonian as the \emph{fast} part. Other choices exist, and order by order the analysis will look different even if a full resummation recovering equivalence of choices were to exist. At any rate, this is effectively the one taken by RD, and we write in our notation their Eq. (B4) accordingly,
\begin{align}
    \epsilon^2H_{\mathrm{fast}}& = \frac{1}{2}\left(p_y^2+ y^2\right)\nonumber\\
    &\quad -\kappa x y + \frac{\kappa^2}{2} x^2+\kappa^2\epsilon_C^2 U_{\mathrm{NL}}\left(\frac{y\Phi_C}{\kappa}\right)\,.
\end{align}
This is a Hamiltonian for the adimensional conjugate pair $(y,p_y)$ in presence of the frozen variable $x$. 

If the conditions for the BO approximation hold, we recover an effective Hamiltonian for the \emph{other} pair, the canonically conjugate pair $(x,p_x)$, by studying the dependence of the ground state energy of the fast Hamiltonian in the variable $x$. Note by the way that RD does not carry out the BO consistency check, but that this is not the brunt of our criticism here. For our purposes at this point, thus, let us assume that BO holds. Then the task at hand is to evaluate $e_0(x;\kappa)$, the ground state energy of $\epsilon^2 H_{\mathrm{fast}}$, and have control over its behaviour as $\kappa\to0$. This defines a Born-Oppenheimer effective potential, 
\begin{align}
    V_{\mathrm{BO}}(x)= \lim_{\kappa\to0}\left[e_0(x;\kappa)-e_0(0;\kappa)\right]\,.
\end{align}
RD expresses this in terms of the dimensionful variable, which explicitly means
\begin{align}
    H_{r,\mathrm{eff}}&= \frac{\kappa^2}{2\epsilon^2}p_x^2 +\frac{1}{\epsilon^2}V_{\mathrm{BO}}(x)\nonumber\\
    &=\frac{Q^2}{2C} +\frac{\hbar}{\sqrt{LC'}}V_{\mathrm{BO}}\left(\frac{\phi}{\Phi_C}\right)\,.
\end{align}
Now, the main technical issue in the computation of $V_{\mathrm{BO}}(x)$ is that  standard Rayleigh--Schr\"odinger perturbation theory cannot be applied directly to $\epsilon^2 H_{\mathrm{fast}}$ with $\kappa$  the perturbation parameter. This comes about because of the presence of $1/\kappa$ in the $U_{\mathrm{NL}}$ term, which impedes the condition of analyticity in the perturbative parameter required for the underlying theorems. RD circumvents this issue in a very elegant manner by using comparison bounds with other Hamiltonians for which the analyticity condition does hold.

RD presents a set of theorems, each applicable to a different class of asymptotic behaviour of $U_{\mathrm{NL}}$. The most salient class, because of the claims of RD that flow from it, is the so-called \emph{sublinear} one, as presented above. The corresponding theorem of RD, Theorem I, states that the effective potential $U_{\mathrm{BO}}=\epsilon^{-2}V_{\mathrm{BO}}$ is zero for sublinear $U_{\mathrm{NL}}$. 

In a way this is not very surprising. Consider an example in the class of sublinear potentials, $U_{\mathrm{NL}}(\phi_c)=-E_J\cos(\phi_c/\Phi_0)$. Thus, the initial quantum Hamiltonian itself has a low-energy purely continuous spectrum. Consequently, the slow Hamiltonian also has a low-energy continuous spectrum, consistent with the result from the theorem, which concerns only the spectral properties and not the scattering data.

\subsection{A first classical analysis}
\label{sec:firstclassical}
As stated repeatedly, RD concludes that Kirchhoff's laws are suspect. They (understandably) do not study \emph{classically} the limit $\kappa\to0$. However, even if the BO approximation is in principle well-known, let us remind the reader that it is associated to adiabatic approximations in the corresponding classical dynamical system. In fact, the intuition from classical systems was a guiding principle in the original work of Born and Oppenheimer \cite{Born:1927}, that explicitly relied on the semiclassical analysis of Born and Heisenberg \cite{Born:1924}. Let us therefore study the classical system under the implicit hypotheses of the previous subsection. That is, we want to study the slow dynamics that follows from the system of conjugate pairs $(x,p_x)$ and $(y,p_y)$ and Hamiltonian $h=\epsilon^2 H_r$, whose classical equations of motion read
\begin{align}
    \dot{x}&=\kappa^2 p_x,\,
    \dot{p}_x= -\kappa(\kappa x-y)\,,\\
    \dot{y}&= p_y,\,
    \dot{p}_y = \kappa x - y-\kappa\epsilon_C^2\Phi_C U'_{\mathrm{NL}}\left(\frac{y\Phi_C}{\kappa}\right)\,. 
\end{align}
To do so we use the slow manifold approach \cite{Fenichel:1979,Jones:1995}, albeit only in very general terms. Namely, we want to identify an approximation of a manifold, invariant under the equations of motion, given by slaving $y$ and $p_y$ to $x$ and $p_x$, that holds for $\kappa\to0$. Inserting $y=\eta(x,p_x)$ and $p_y=\pi(x,p_x)$, we obtain consistency equations for the two functions $\eta$ and $\pi$. These are partial differential equations, and were we to obtain their general solution we would have a general family of two-dimensional manifolds invariant under the dynamics. However, we are interested in the \emph{slow} manifold. In many situations this is identified perturbatively. Here there can be an obstacle to a perturbative expansion, in parallel with the applicability or otherwise of analytic perturbation theory for the fast Hamiltonian above. Namely that the small parameter $\kappa$ enters non-analytically in the consistency equation. Explicitly, the system of partial differential equations under study is
\begin{align}\label{eq:invariantsystem}
    \pi &= \kappa\eta \frac{\partial\eta}{\partial {p_x}} +\kappa^2\left[p_x \frac{\partial\eta}{\partial {x}} - x \frac{\partial\eta}{\partial {p_x}}\right]\,,\\
    \eta &= \kappa\left[x-\epsilon_C^2\Phi_C U'_{\mathrm{NL}}\left(\frac{\Phi_c\eta}{\kappa}\right)\right]\\
    &\quad -\kappa\eta \frac{\partial\eta}{\partial {p_x}}-\kappa^2\left[p_x \frac{\partial\eta}{\partial {x}} - x \frac{\partial\eta}{\partial {p_x}}\right]\,.\nonumber
\end{align}
Since the example we always have in mind is that $U_{\mathrm{NL}}(\phi_c)=-E_J\cos\left(\phi_c/\Phi_0\right)$, let us make the assumption that $U'_{\mathrm{NL}}$ and all its derivatives are bounded from above and from below. In that case we can indeed look for perturbative solutions to the system \eqref{eq:invariantsystem}. By inspection, $\eta=O(\kappa)$ and $\pi=O(\kappa^3)$. The leading term of $\eta$, denoted by $\eta_1$, must then be the solution to the (possibly transcendental) equation
\begin{align}\label{eq:sloweta1}
    x = \eta_ 1+\epsilon_C^2 \Phi_C U'_{\mathrm{NL}}\left(\Phi_C\eta_1\right)\,.
\end{align}
This equation is exactly the consistency equation for the circuit without the parasitic capacitance, earlier presented in Eq. \eqref{eq:betaeq}, and controlled by the $\beta$ parameter of Eq. \eqref{eq:betadef}. Thus, \emph{if} this analysis is consistent, we recover, as expected, the simpler circuit as the adiabatic slow manifold approximation of the full circuit. Note, however, that the validity of this adiabatic approximation requires several conditions. In particular, uniqueness for $\eta_1$ is conditional of $1+\epsilon_C^2\Phi_C^2U''(\Phi_C\eta)>0$ for all $\eta$. If any of these conditions are not met it is indeed the case that the removal of the parasitic capacitance is a \emph{singular} operation, even classically. 

In the region of validity of this approximation, we still have to compute the effective potential. This is defined parametrically by Eq. \eqref{eq:sloweta1} together with
\begin{align}
    V'(x)&= \epsilon^2_C\Phi_C U'_{\mathrm{NL}}\left(\Phi_C\eta_1\right)\nonumber\\&= x-\eta_1\,.
\end{align}
For this case one can actually compute $V(x)$ directly in terms of $\eta_1$, and one has
\begin{align}
    V(x)= \epsilon_C^2 U_{\mathrm{NL}}(\Phi_C\eta_1) +\frac{1}{2}\epsilon_C^4\Phi_C^2 \left[U_{\mathrm{NL}}'\left(\Phi_C\eta_1\right)\right]^2\,.
\end{align}
\subsection{First conclusion}
Let us consider explicitly the case of  $U_{\mathrm{NL}}(\phi_c)$  being infinitely differentiable on the real line and with it and all its derivatives being of bounded variation on the line. This is a subcase of sublinear potentials in the classification of RD, and thus the prediction of RD is that the effective Hamiltonian for the low-energy sector is zero. This seems to be in contradiction with the classical computation of the previous subsection, where (under some conditions) we explicitly compute a non-zero effective potential $V(x)$, and thus appears to lend weight to their conclusion that something is wrong with Kirchhoff's laws. 

In fact, that is not the situation at all. Take the by now canonical example   $U_{\mathrm{NL}}(\phi_c)=-E_J\cos(2\pi\phi_c/\Phi_Q)$, and let $\beta < 1$. Then the computation above shows that $V(x)$ leads to a low-energy purely continuous spectrum. This makes the classical adiabatic limit and then quantization fully consistent with quantization and then BO approximation. 

Bear in mind that the classical analysis \emph{assumes} the validity of Kirchhoff laws for the full circuit. But, more to the point, the slow manifold adiabatic approximation provides us with a condition that is an expression of current conservation for the reduced circuit.  And under the implicit assumption that $\phi_c$ and $Q_c$ are variables in the real line, there is complete consistency with the results of RD.

Furthermore, there is no failure of Dirac-Bergmann: the adiabatic condition is actually the same as the Dirac-Bergmann (or Faddeev--Jackiw) constraint. The ``breakdown" of Dirac-Bergmann is in the same range for which the simple adiabatic approximation is not valid.

Thus, even setting aside the need to consider compact variables, as we shall in the next section, neither the conclusions drawn in RD from their analysis in Sec. II nor those derived from Thm. 1 are supported.

\section{Does Theorem 1 apply if $\phi_c\in S^1$?}
\label{sec:No_Th1_S1}
We have already indicated that unstated hypotheses are required for the previous analysis. Most importantly in the context of superconducting circuits, the classification of nonlinear potentials itself only makes sense if the domain of the potential is the full real line. This assumption, that the variables are supported on the full real line, is implicit throughout the analysis of RD. We shall now argue that, setting that assumption aside,  the analysis of RD is inapplicable when $\phi_c$ is a compact variable. 

Since there seems to be some confusion in the literature regarding compact variables and the periodicity of potentials, we will present several aspects of the central argument to shed light on the difference. The initial set of perspectives will be classical, and the quantum mechanical side will be presented in the last subsection.
\subsection{No asymptotic classification}
There is an almost trivial observation to be made first: the classification of potentials in three classes that is central to the  theorems in RD does not carry through for compact $\phi_c$. To be more precise, the classification does not carry through for nonlinear potentials whose domain is a finite range. Thus, sublinear, superlinear or linear are all merged. As the results in the corresponding theorems for the full real line depend essentially on that classification, it is almost a truism that they are not informative nor relevant for the compact case.
\subsection{Intrinsic scale}

As stated above, the adimensionalization of $\phi_c$ and $Q_c$ presented above seems uncontroversial. After all, units are simply conventional. However, that changes if there are \emph{intrinsic scales} in the problem, as then we have additional parameters to contend with. Such a case is presented if a variable lives on the circle. In the case of interest to us, $\phi_c$, associated with the Josephson junction, presents with a natural flux scale, the magnetic flux quantum $\Phi_Q$ . This flux scale is dual to a charge scale, $2e$, the charge of a Cooper pair.   

Let us now return to $H_r$, Eq. \eqref{eq:Hr_RD_circuit}, and adimensionalize first with $\phi_c=\Phi_Q \varphi_c/2\pi$ and $Q_c= 2 e n_c$. We have 
\begin{align}
    \mathrm{d}Q_c\wedge\mathrm{d}\phi_c = \hbar \mathrm{d}n_c\wedge\mathrm{d}\varphi_c\,.
\end{align}
Similarly with $\phi$ and $Q$, and introducing $E_C=4e^2/C=\kappa^4 E_C'$, as well as a nonlinear energy scale $E_U$  and an adimensional function $u(\varphi)$ such that $U_{\mathrm{NL}}(\phi_c)= E_U u(\varphi_c)$, we have
\begin{align}\label{eq:hradim2}
    H_r&= E_C'\left[n_c^2\right.\nonumber\\
    &\quad\left.+\kappa^4 n^2 +\frac{1}{2}\kappa^4\left(\frac{\hbar\omega_C}{E_C}\right)^2\left(\varphi-\varphi_c\right)^2+\kappa^4\frac{E_U}{E_C}u(\varphi_c)\right]\,.
\end{align}
Crucially, this is exactly the same Hamiltonian as in Eq. \eqref{eq:adimHr}, but it seems that the hierarchy fast/slow is skewed. That is not necessarily the case. In all the coming discussion  $n$ and $\varphi$ are assumed to  be taking values in the real line. Consider first $\varphi_c$ also an extended variable (full real line). Then were we to take $n_c^2$ as the fast Hamiltonian we would run into trouble, both classically and quantum mechanically. From a quantum mechanics perspective, as the spectrum of $n_c$ is the whole real line in this situation, there is no adiabatic separation and BO cannot be carried out consistently with this choice of fast Hamiltonian. 

However, if $\varphi_c$ is restricted to the circle, the spectrum of $n_c$ is $\mathbb{Z}+n_g$, with $n_g\in[0,1)$ the gate charge bias parameter, and $n_c^2$ can be chosen as a fast variable. Here we see a crucial dissimilarity between the two relevant cases.

At this point, a question can arise regarding the case of the full real line. How can it be that both presentations of $H_r$ provide us with the same adiabatic structure as $\kappa\to0$ in this situation? This point will be addressed next.  

\subsection{Dilations and scales (classical mechanics)} \label{sec:dilclass}

We need a more critical look at what changes of units really are in this Hamiltonian context: dilation transformations. Thus we look at both classical and quantum mechanical dilations, with special emphasis on two aspects. First, that in the limit $\kappa\to0$ the transformations are singular and, second, that there are no canonical dilation transformations for $S^1$ configuration spaces.

Consider the vector field
\begin{align}
    X_D= p_y\frac{\partial}{\partial p_y}-y\frac{\partial}{\partial y}
\end{align}
in a patch with coordinates $y$ and $p_y$. 
The coordinate functions $y$ and $p_y$ obey $X_D(y)=-y$ and $X_D(p_y)=p_y$. Therefore there is a group of local transformations (dilations) of the form $\exp(\lambda X_D)$, such that $e^{\lambda X_D}y= e^{-\lambda}y=y(\lambda)$  and $e^{\lambda X_D}p_y= e^{\lambda}p_y= p_y(\lambda)$. Furthermore, locally we have 
\begin{align}
    \mathrm{d}p_y\wedge\mathrm{d}y=\mathrm{d}p_y(\lambda)\wedge\mathrm{d}y(\lambda)\,,
\end{align}
so they seem to be canonical transformations. If the symplectic manifold (dually Poisson manifold) is $\mathbb{R}^2\simeq T^*\mathbb{R}$ then this transformation is indeed canonical, as the coordinate patch covers the full manifold. 

Now, in most cases when a change of   adimensionalization of variables is carried out in the context of  classical mechanics there is an implicit assumption: that equivalent dynamics is obtained in both descriptions. As the change of variables is time independent, the requirement is that the transformation be symplectic (canonical), i.e. that it preserve the symplectic form (Poisson bracket). 

Under the assumption that  the underlying manifold is $\mathbb{R}^4\simeq T^*\mathbb{R}^2$, Hamiltonians \eqref{eq:adimHr} and \eqref{eq:hradim2} are connected through dilations (scale transforms) with two parameters,
\begin{align}\label{eq:scalewithkappa}
    \lambda_y &= \ln \frac{\Phi_Q}{2\pi\Phi_{\mathrm{ZPF}}}= \lambda_x +\ln\kappa\,,\\
    \lambda_x &= \ln \frac{\hbar}{2e\Phi_{C}}\,,\nonumber
\end{align}
the transformation is canonical, and full equivalence is recovered. 
Observe that the parameter $\lambda_x$ is independent of $\kappa$, and thus that $\lambda_y$ is singular as $\kappa\to0$.

Amongst other properties, this shows that the determination of  the slow Hamiltonian is not achieved simply through reading out the coefficients of various terms of the full Hamiltonian. Rather, for any choice of equivalent Hamiltonian one must make ans\"atze and, a posteriori, check whether the adiabatic approximation holds. Alternatively, one searches for one equivalent presentation of the Hamiltonian in which indeed the adiabatic separation is more intuitively presented. We stress that this is not a purely quantum mechanical artefact. It is in fact a central consideration whenever a slow-fast separation is being investigated in dynamical systems.

 \subsection{Scale changes and limits}
 There is another issue when considering limits of a parameter in a Hamiltonian in combination with (even canonical) scale changes, that is best understood by the example of a harmonic oscillator, with Hamiltonian $p^2/2m+kx^2/2$. To adimensionalize, we choose a length scale $\ell$ and a time scale $\tau$, such that $x=\ell \xi$ and $p=m\ell \pi/\tau$.  We also introduce the adimensional $\kappa= k\tau^2/m$, to write the Hamiltonian as $H= (m\ell^2/\tau^2)\left(\pi^2+\kappa \xi^2\right)/2$, with canonical $\left\{\xi,\pi\right\}=1$. A canonical transformation $\pi\to e^\lambda \pi$ and $\xi\to e^{-\lambda} \xi$ provides us with an equivalent Hamiltonian $H_\lambda =(m\ell^2 e^{2\lambda}/\tau^2)\left(\pi^2+\kappa e^{-4\lambda} \xi^2\right)$. By choosing $\lambda=(1/4)\ln\kappa$ we have the canonical form $H_{\mathrm{can}}=\left(m\ell^2/\tau\right)\omega\left(\pi^2+\xi²\right)/2$, with $\omega=\sqrt{k/m}$, as is well known. Let us now consider the limit $k\to0$, for fixed mass. The limits of the initial Hamiltonian and of the canonical one are completely different, as in the limit $k\to0$ the initial Hamiltonian tends to the free particle, while the canonical one tends to a nondynamical system. The reason is that the  parameter $\lambda$ required to reach the canonical form tends to $-\infty$. The change of scales transformation that connected the two equivalent Hamiltonians becomes singular. 

 The conclusion of this argument is that both classical and quantum adiabatic limits have to be carefully considered in families of Hamiltonians which are connected by transformations that become singular as the small controlling parameter tends to zero.  In the case of interest, we have shown in Eq. \eqref{eq:scalewithkappa} that scale change is divergent with $\kappa\to0$, thus demonstrating that in that limit the equivalence between Hamiltonians \eqref{eq:adimHr} and \eqref{eq:hradim2} is suspect.
 
 \subsection{Non-symplectic classical dilations on the circle}

Additionally, and crucially for our purposes, the dilation group is \emph{not} symplectic (canonical) for an $S^1$ configuration space.  The full dilation group generated by $X_D$ for $\mathbb{R}^2\simeq T^*\mathbb{R}$ breaks down with the compactification of the configuration space. Dilations are \emph{locally} canonical, but not globally so. 

Let us  present the argument using the Poisson structure. The starting point is $\mathbb{R}^2$ with the natural Poisson bracket. With conjugate coordinates $a$ and $z$, this means 
that for $f,g\in C^\infty(\mathbb{R}^2)$
\begin{align}
    \left\{f,g\right\}=\frac{\partial f}{\partial a}\frac{\partial g}{\partial z}-\frac{\partial f}{\partial z}\frac{\partial g}{\partial a}\,.
\end{align}
The generator of dilations issues through the Poisson bracket from the function $D(a,z)= a z$, whereby we mean that
\begin{align}
    X_D(f)=\left\{D,f\right\}\,.
\end{align}

The Poisson structure of the cylinder can be obtained by quotienting $\mathbb{R}^2$, with conjugate coordinates $a$ and $z$, in the $a$ direction, by identifying $\left(a+2\pi,z\right)\sim\left(a,z\right)$ for all $a$ and $z$. Let us do this explicitly. The coordinate function $z$ on $\mathbb{R}^2$ provides us with a vector field $X_Z$, such that for all $C^\infty(\mathbb{R}^2)$ functions $f$ we have
\begin{align}
    X_Z(f)=\left\{z,f\right\}=-\frac{\partial f}{\partial a}\,.
\end{align}
This vector field is the generator (acting on functions) of translations along the $a$ direction, 
\begin{align}
    \left[e^{\mu X_Z}f\right](a,z)= f(a-\mu,z)\,.
\end{align}
Let
\begin{align}
    T=\exp\left(2\pi X_Z\right)\,.
\end{align}
This is the discrete translation by $2\pi$ along the $a$ direction, acting on functions. This discrete translation generates a $\mathbb{Z}$ group, with $\mathbb{Z}\ni n\rightarrow T^n$. Crucially, $T$ (and consequently $T^n$) is a Poisson map. I.e. for all $f,g\in C^\infty(\mathbb{R}^2)$, we have
\begin{align}
    \left\{T(f),T(g)\right\}=T\left(\left\{f,g\right\}\right)\,.
\end{align}
Thus the cylinder as a symplectic manifold is obtained by the identification above and the restriction to $C^\infty(\mathbb{R}\times S^1)$, with the induced Poisson bracket being the natural Poisson bracket on the cylinder.

Turning back to our objective, $T$ and $X_D$, both acting on $C^\infty(\mathbb{R}^2)$, do not commute,
\begin{align}
    \left[T,X_D\right]= 2\pi X_a T\,,
\end{align}
with $X_a=\partial_a$. Therefore, it follows that $X_D$ takes us out of the cylinder, and that it cannot be a Poisson map. In other words, even if the change of scale (dilation) is \emph{locally} symplectic, it is not globally symplectic for the cylinder. 

Alternatively, it maps from one cylinder to another, different, cylinder, changing its radius. 

Combining this analysis with the issue of singular limits, we see that  the two Hamiltonians at play, Eqs. \eqref{eq:adimHr}  and \eqref{eq:hradim2}, are not equivalent in the limit $\kappa\to0$ in the case of compact configuration space.
\subsection{Dilations and scales (quantum mechanics)}
 The argument above has a quantum mechanical correlate, with some twists. Translations along the (adimensional) configuration space $\mathbb{R}$ are realized classically on functions as above, with generator $X_a$. The quantum mechanical version is the unitary group $\exp\left(\mu \partial_a\right)$  , with generator $\hat{p}_a=-i\partial_a$, as $U(\mu)=\exp\left(i \mu\hat{p}_a\right)$.

 As to dilations, the symplectic transformation on $\mathbb{R}^2$ is mapped to a unitary transformation, with generator $\hat{D}=\frac{1}{2}\left\{\hat{a},\hat{z}\right\}$ for canonical $\left[\hat{a},\hat{z}\right]=i$. It is again the case that $\left[\hat{z},\hat{D}\right]\neq0$, and we again see that there is no unitary that acts a scale change for functions on an interval $(0,2\pi)$. 

 When we speak about quantization on the circle, what we actually mean is on an interval with the additional concept of translation invariance, which is geometrically possible on the circle by identification of the boundary.  Thus, the issue is the identification of a self-adjoint generator of translations. The general theory of self-adjoint extensions allows the identification of the $U(1)$ family of self-adjoint generators of translations on the circle of radius 1, $\hat{n}_\gamma$, with spectrum $\sigma(\hat{n}_\gamma)=\mathbb{Z}+\gamma$. Here $\gamma\in[0,1)$ determines a monodromy $\exp(2\pi i\gamma)\in U(1)$. The canonical commutation relations are obeyed in that on its domain $\hat{n}_\gamma=-i\partial_\varphi$, with $\varphi$ the variable in the interval $(0,2\pi)$, for each $\gamma$. The parameter distinguishing the generator does not appear in the differential expression, $-i\partial_\varphi$, identical for all of them. Rather, it identifies the domain such that differential expression becomes a bona fides self-adjoint operator. 

 At any rate, whichever generator of translations one considers, a hypothesized scale transformation cannot be implemented by a unitary: essentially, a transformation $\hat{n}_\gamma\to e^\lambda \hat{n}_\gamma$ would change the spectrum, and spectra are invariant under unitary conjugation. In the $\mathbb{R}$ case, on the other hand, the spectrum of the generator of translations is the full real line. The real line is invariant under the transformation $x\to e^\lambda x$, and indeed  scale changes are implemented by unitaries.

 \section{Adiabatic limit for $\phi_c\in S^1$}
\label{sec:Adiabatic_S1}
 In light of the previous section, we now examine the low energy sector of Hamiltonian \eqref{eq:hradim2} when $\varphi_c$ is a coordinate on the circle of radius 1. We set aside the global energy scale $E'_C$ in what follows. For definiteness, we start with the assumption that $\varphi$ takes values in the full real line.
 \subsection{Classical slow manifold approximation}
 We start with Hamiltonian \eqref{eq:hradim2}, with $E'_C$ set to one (equivalently, we choose a suitable time unit). We now make the slow manifold approximation $\varphi_c=\varphi_c(\varphi, n)$ and $n_c=n_c(\varphi,n)$, with a slight abuse of notation. We arrive at the following consistency equations for the slow manifold:
 \begin{subequations}\label{eq:slowalternative}
 \begin{align}
     n_c &= \kappa^4\left[n\frac{\partial \varphi_c}{\partial\varphi}+\frac{1}{2}\left(\frac{\hbar\omega_C}{E_C}\right)^2\left(\varphi_c-\varphi\right)\frac{\partial\varphi_c}{\partial n}\right]\,,\label{eq:slownc}\\
     \varphi &= \varphi_c+\left(\frac{E_C}{\hbar\omega_C}\right)^2 \frac{E_U}{E_C}u'(\varphi_c)\label{eq:slowcomplicatec}\\
     &\qquad + 2\left(\frac{E_C}{\hbar\omega_C}\right)^2 n\frac{\partial n_c}{\partial \varphi}+ \left(\frac{E_C}{\hbar\omega_C}\right)^2 \frac{\hbar\omega_C}{E_C}\left(\varphi-\varphi_c\right)\frac{\partial n_c}{\partial n}\,.\nonumber
 \end{align}
 \end{subequations}
 By inspection, we see that we can consistently demand  $n_c=O(\kappa^4)$ and $\varphi_c=O(1)$. Thus, to lowest order, we have the approximate consistency equation
 \begin{align}\label{eq:phiphic}
     \varphi = \varphi_c+\frac{E_C E_U}{\hbar^2\omega_C^2}u'(\varphi_c)\,,
 \end{align}
 which for the standard example is nothing but Eq. \eqref{eq:betaeq}, yet again. We recover the classical effective Hamiltonian for $n$ and $\varphi$ as above, again for $\beta<1$ in the example. 

To better appreciate this result, we should point that the classical equations of motion are local and do not by themselves detect the global structure of the underlying phase space. Nonetheless, this global structure crops up again in the determination of the slow manifold approximation. Let us restrict ourselves again to the case $1+E_CE_U u''(\varphi_c)/\hbar^2\omega_C^2>0$ for all $\varphi_c\in[0,2\pi]$, ensuring invertibility. Then $\varphi$ can only range from 0 to $2\pi$, and the effective potential
\begin{align}
    V(\varphi)= \frac{E_U}{E_C}\left\{u(\varphi_c)+\frac{1}{2}\frac{E_C E_U}{\hbar^2\omega_C^2}\left[u'(\varphi_c)\right]^2\right\}
\end{align}
is periodic in $\varphi$ with period $2\pi$, again, see how this perfectly matches the analyses in the supplementary  material/information of \cite{Smith:2023} and \cite{Willsch:2024}. Thus the effective Hamiltonian will have a discrete spectrum.

One can additionally check that, under the invertibility condition,  $V(\varphi)$ will present as many minima as $u(\varphi_c)$ does, since $V'(\varphi)=(E_U/E_C)u'(\varphi_c)$. The frequency of the harmonic approximation for a minimum of $u(\varphi_c)$ will be modified for the corresponding minimum of $V(\varphi)$, since 
\begin{align}
    V''(\varphi)= \frac{E_U}{E_C}\frac{u''(\varphi_c)}{1+E_C E_U u''(\varphi_c)/\hbar^2\omega_C^2}\,.
\end{align}
For the example $U_{\mathrm{NL}}(\phi_c)=-E_J\cos(2\pi\phi_c/\Phi_Q)$, i.e. for $E_U=E_J$ and $u(\varphi_c)=-\cos(\varphi_c)$, the single minimum is at $\varphi_c=0$ (that is identified with $2\pi$).
\subsection{Na\"{\i}ve quantum adiabatic limit}
 Subtleties would arise were the gate charge such that $|n_g-1/2|\leq\kappa^2$, so we look at the more general case and  consider the gate charge to be zero. The fast Hamiltonian from Eq. \eqref{eq:hradim2} is thus $n_c^2$, with spectrum $0,1,4,...$, independent of $\varphi$. The ground state of this fast Hamiltonian has the wavefunction $\psi_0(\varphi_c)=1/\sqrt{2\pi}$. On computing the (operator valued) expectation value of the slow Hamiltonian in this state, we are led to an effective Hamiltonian that is a displaced harmonic oscillator, and its spectrum is $e_{\mathrm{slow},k}= \sqrt{2}\kappa^4\left(\hbar\omega_C/E_C\right)\left(k+1/2\right)$, with $k\in\mathbb{N}$. 

 There is an equivalent argument found in the literature, as follows: $n_c$ must be fixed, and therefore $\varphi_c$ must be completely undetermined. Thus the effective Hamiltonian is obtained by averaging over the circle. 

 We see that there is a contradiction in this case between the quantization of the classical adiabatic approximation and this last argument. Setting minor details aside, this is palpable in that the classical adiabatic approximation includes the ratio $E_U/E_C$ as a relevant parameter, while it is absent in the displaced harmonic oscillator limit. This indicates that the limits $L\to0$ and $C'\to0$ do not seem to commute, if one takes the quantum adiabatic limit as in the paragraph above. 
 
 Both positions (classical elimination/adiabatic limit and then quantization, on the one hand, and quantization and then BO of some form) have been taken in the literature in the presence of compact variables, and we offer here no solution to the conundrum. 

 Ironically, one of the elements of RD was the perceived contradiction between the quantization after imposing reduction with equation \eqref{eq:betaeq} (or its generalizations) and the BO computation. Yet, as has been shown above, they are both compatible under the assumptions implicit in RD for the relevant case. However, the distinction with $S^1$ variables was set aside in RD, and prima facie there is a contradiction between quantization after adiabatic reduction and direct BO in this situation.

\section{Lack of experimental evidence}
\label{sec:Experimental_evidence}
In this section, we show  that the main conclusions of RD are not supported by experimental evidence. First, we briefly review the standard blackbox modeling of linear environments in superconducting circuits, with increasingly complex lumped-element circuits, and then indicate the consequences that the claims in RD would have. Second, we discuss  experimental efforts where parts of their superconducting circuits have been modeled with the circuit in Fig.~\ref{fig:RD_Circuit}(c), or versions thereof, in the nonsingular parameter regime, with successful matching to the experimental results.

\subsection{Lumped-element blackbox modeling of linear electromagnetic environments}

The task of  modeling real superconducting circuits, which are very complex three-dimensional (3D) systems, has been carried out in the main, and by now for several decades, by  using phenomenological macroscopic classical lumped (or quasi-lumped~\cite{ParraRodriguez:2018,Minev:2021,ParraRodriguez:2024}) circuit descriptions as the starting point. Once a well-behaved classical theory has been constructed, the rules of canonical quantization have been imposed on those macroscopic degrees of freedom~\footnote{Originally, the path integral formulation of quantum mechanics was rather preferred to model dissipative effects in superconducting circuits behaving in the quantum regime, e.g.,  ~\cite{CaldeiraLeggett:1981}.}. 

Naturally, depicting 3D superconducting matter, and its environment, in terms of (0D) lumped elements  is a drastic elimination of scales. However, as  happens with circuits  in the classical regime, this approximation works extremely well for signals  above a certain wavelength, with threshold  typically controlled, as to order of magnitude, by the length of the circuit~\cite{Jackson:1999,Pozar:2009}. In the context of  superconducting circuits the following strategy has been put to good use. Namely, to split the description of the device into its linear and nonlinear behaviour, the latter typically  due to  the Josephson effect~\cite{Josephson:1962}. Subsequently, one makes use of a linear response function (generally a scattering matrix $\mathsf{S}(\omega)$, or, particularly,  impedance $\mathsf{Z}(\omega)$ or admittance $\mathsf{Y}(\omega)$ matrices), \emph{seen} from what would be small pairs of terminals defining the Josephson junctions, that fits/simulates the  classical behaviour of the linear part {to a desired level of accuracy}. Expansion theorems~\cite{Foster:1924,Brune:1931,Newcomb:1966,Anderson:1975,Solgun:2015} for linear response functions are used to construct lumped~element \footnote{More generally, quasi-lumped models.}  circuit models that match this behaviour in a wide enough range of frequencies. Finally, the lumped-element model of the JJs is added in the ports of the circuit representing the linear environment, and a canonically quantized Hamiltonian is derived. This approach was coined as \emph{blackbox quantization} when applied to systems consisting of Josephson junctions suspended in large 3D cavities~\cite{Nigg:2012, Solgun:2014, Solgun:2015}, although it is applicable and has been implicitly used before in all previous modeling of planar circuits as well.

\begin{figure}[!ht]
	\includegraphics[width=1
\linewidth]{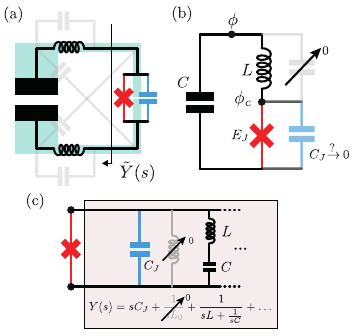}
\caption{(a) A conceptual sketch layout of a Josephson junction (small gap) shunted by a large capacitance (big gap) can have increasingly precise lumped element models accounting for various capacitive and inductive effects. In a \emph{first-order} model, the response of the shunting linear environment is often approximated (e.g., \cite{Koch:2007} and subsequent literature) by only the pole at $s=-i\infty$ of the admittance function $\tilde{Y}(s)$. (b) In a \emph{second-order} approximation, the inductive response of the superconducting loop can be added, reaching the motivating circuit used in RD. Note that here, and in the circuits of Figs. 3 and 4 of RD, the capacitive effects in parallel with the inductance $L$ are classically neglected before any quantum analysis is performed~\cite{Rymarz:2018}. (c) More systematically, the linear response of a one-port lossless environment can be modeled to the desired level of precision with a series sum of poles at infinity ($C_J$), zero ($L_0$), and finite frequencies (here, a first pole at $\omega=1/\sqrt{LC}$) \cite{Newcomb:1966,Nigg:2012,Solgun:2014,Solgun:2015}. Note that, from the purely mathematical blackbox perspective, not all the nodes of a lumped-element circuit must be connected through capacitors, nor do all the connecting wires have an associated inductance.
}
\label{fig:Physically_motivated_lumped_element_motivated}
\end{figure}

This strategy can  be applied to model a realistic physical realization with  circuits of the family of Fig.~\ref{fig:RD_Circuit}, according to the conceptual circuit layout in Fig.~\ref{fig:Physically_motivated_lumped_element_motivated}(a). This presents two superconducting islands (light green pieces), in a twice broken loop. The blackbox approach  considers the small gap, where a considerable amount of quantum tunneling behaviour would be observable, as the nonlinear element (JJ), and the rest of the circuit as a linear environment. A rough lumped element modeling of the circuit would require splitting the total area in several subparts, and map each subpart to a lumped element. In this case,  the black circuit overlayed on the device sketch  includes a capacitor in series with two inductances, and a parasitic capacitance in parallel with the junction port. Naturally, in this first approximation, we would be neglecting, among other things, parasitic capacitive effects between other points (light grey capacitors). With standard reductions (in turn making use of Kirchhoff's laws) we arrive at Fig.~\ref{fig:Physically_motivated_lumped_element_motivated}(b), where, again, all possible capacitive effect in parallel with the inductance is deemed irrelevant, as mentioned by one the authors in~\cite{Rymarz:2018}, and assumed in RD.

A more systematic application of the blackbox approach would be to probe the port defined by the junction, and simulate/fit the total response, for instance the admittance function $y(t)$. If no dissipation were considered, by means of Foster's theorem~\cite{Foster:1924}, its response would have leading order coefficients for the poles at infinity ($C_J$), and at a finite frequency $\omega=1/\sqrt{LC}$. In other words,  its response (in Laplace space) is fit to the complex function
\begin{align}
    Y(s)=C_J s + \frac{s/L}{s^2+ \Omega^2}+\dots\, .
\end{align}
The infinite and zero poles of the response functions are mathematically possible from the causal point of view.  For a wide frequency range they provide a very good approximation. In reality, however, such poles would always be at finite frequencies. In many experiments, where the junction and the big capacitor are much closer (such as the interdigitated capacitor used in transmon qubits~\cite{Houck:2009}), inductive effects of the kind modelled by $L$  in Fig.~\ref{fig:Physically_motivated_lumped_element_motivated}  are typically neglected $L\rightarrow 0$ (similarly in Fig.~\ref{fig:RD_Circuit}), and researchers  fit data to the total parallel capacitance $C_{\parallel}=C+C_J$.

Let us  reevaluate some of the assumptions and implications of RD's analysis regarding the application of Thm 1 to bounded nonlinear potentials, now {in} the context of this discussion. First of all, if Thm 1 were applicable to the circuit in Fig.~\ref{fig:Physically_motivated_lumped_element_motivated}(b,c), its result would mean that a parasitic value of $L$ (say, $\delta L$) would appear to dramatically change the approximate low-energy Hamiltonian model. That is, on the one hand (no inductance considered, $L=0$), we would have a standard CPB-transmon Hamiltonian
\begin{align}
    H&=\frac{Q^2}{2(C+C_J)}-E_J\cos(2\pi \phi/\Phi_Q),\label{eq:H_transmon_C_CJ}\\
    &\approx\frac{Q^2}{2C}-E_J\cos(2\pi \phi/\Phi_Q)\label{eq:H_transmon_C}\,,
\end{align}
where, in the absence of $L$, $\phi\equiv \phi_c$, whereas in the other RD predicts (independent of the value of $L$) 
\begin{align}
    H_{r,\text{eff}}\approx \frac{Q^2}{2 C}\,.\label{eq:H_transmon_Th1}
\end{align}
 Two very different results seem to arise, as a consequence of taking different limit orderings (first $L \rightarrow 0$, or first $C_J \rightarrow 0$). As indicated above, if both flux and charge take values over the whole real line, as was assumed implicitly in RD, these two models both present identical low-energy purely continuous spectrum.
 
 Yet, and contrary to the analysis of RD, the  models for the transmon with Hamiltonians  (\ref{eq:H_transmon_C}) or (\ref{eq:H_transmon_C_CJ}) seem to be in good agreement with experiments when the flux coordinate is considered a compact variable (and thus its conjugate charge presenting a discrete spectrum). Furthermore, and in line with all the previous arguments, we note the discrepancy between the conclusions derived in RD from Thms 1 and 2 and the effective nonlinear potentials for nonsingular screening parameters found by Miano et al.~\cite{Miano:2023}.

\subsection{Nonsingular circuit models without parasitic capacitances}

Beyond the previous discussion regarding transmon-type qubits, there has been extensive experimental modeling of superconducting circuits where the parasitic capacitance of the Josephson junction has been removed prior to the quantization of the Hamiltonian in the nonsingular regime. For instance, as early as \cite{Ilichev:2001} (see also the discussion in Ref.~\cite{Golubov:2004}), the authors constructed a circuit containing an inductor  of significantly larger inductance than the parasitic one of the connecting wire, in series with a number of parallel Josephson junctions whose current-flux relation at cryogenic temperatures was well described by an effective nonlinear inductor, in the spirit of the nonlinear potential of Eq.~(\ref{eq:H_s_RD}), for the JJ, and $\beta<1$. This effective model is a generalization of the problem sketched in Fig.~\ref{fig:RD_Circuit}(c) (Fig. 3 in Ref.~\cite{Rymarz:2023}), with generalized $\beta_i$ parameters {for the set of JJs}.

Several other works have used the nonsingular parameter regime of $\beta < 1$ to model higher Josephson harmonics. For instance, examples of the application of the circuit model in the $\beta \ll 1$ regime concern the so-called Kinetic Interference coTunneling Elements (KITEs)~\cite{Smith:2020, Bell:2014}. A circuit consisting of two nonlinear arms, where each of them is trivially approximated to first order (neglecting all parasitic capacitances) by an inductor in series with a pure JJ element~\cite{Smith:2023}. Most recently, Ref.~\cite{Willsch:2024} (see Supplementary Information) even discusses the possibility of the nonsingular model ($\beta \ll 1$) as a possible explanation for the corrections to the standard cosine function for Josephson's potential. Interestingly, and in contrast to the analysis in RD, there  the node flux variable $\phi$ was argued to be considered compact, as the inductance is indeed in series and not in parallel with the junction. Incidentally,  this  clearly demonstrates that the debate between extended and compact modeling of the phases is far from  being settled.

At any rate, we note that  nonsingular ``$\beta$-models" seem to be in agreement with  experimental results, and in strong disagreement with the conclusions in RD. We mention in passing that the multivaluedness in Eq. \eqref{eq:sloweta1} when $\beta>1$ (singular $\beta$-model) has been used to address hysteretic behaviour and heating in Josephson junctions~\cite{Gumus:2023}. We do not investigate here the possible significance of that perspective for effective quantum circuital descriptions. 

All things considered, we believe that there is sufficient experimental evidence to argue that the na\"{\i}ve application of Theorems 1 and 2 to bounded even potentials based on (possibly parallel connections of) Josephson junctions is entirely unjustified.

\section{Summary}
\label{sec:Summary}
In this summary section, we first give a discursive summary of our understanding of Rymarz' and DiVincenzo's argument in RD, followed by an equally discursive summary of our objections. As both argument and criticism are unfortunately rather involved, this is followed by a point-by-point  summary following the order in RD. We advise the reader that an even more compact version of this discussion is presented in Fig.~\ref{fig:Summary_criticism}. We add some concluding remarks.

\subsection{Discursive summary}
The discussion in the last section reveals  that in the construction of circuital models of superconducting devices there are many ambiguities. Yet, because of predictive power for experiments it is the case that for some range of parameters simplifications are justified, in particular with elimination of parasitic immitances. RD is of course correct in reminding us that for other ranges outright elimination of parasitic elements might lead to trouble. However, both their argument to that point and their framing of the assertion  are flawed. Given this strong statement of ours, we ought to make explicit what is it that we understand as their argument.  

Centrally, it goes as follows: we examine some families of circuits that are singular from the node-flux perspective. Using some Hamiltonian reduction method (they use in particular Dirac-Bergmann) we identify constraints that are required for consistency. It is to be noted that these constraints are associated with Kirchhoff's laws. We then obtain reduced Hamiltonian systems, but there are critical points in parameter space, separating pathological and non-pathological regions. RD assert that this is a failure of the Dirac--Bergmann algorithm. Whatever the origin, whenever these critical points appear  we should look for regularizations, in particular parasitic immitances. As the quantization of Hamiltonians in which those immitances appear is uncontroversial, it behooves us to study the quantum mechanics of the limit when the parasitic elements are smoothly removed. This is carried out through the Born-Oppenheimer method, applicable when there are strongly separated energy scales.

This part of the argument could be understood as ``going down", i.e., as identifying a computation whence to draw conclusions from. We have no major issue with the perspective in RD to this point, other than RD assigning blame to the Dirac--Bergmann method (but see below).

Next, ``going up", RD chooses a particular slow-fast separation, and their fast Hamiltonian includes the small parameter. As it does not enter analytically, they demonstrate technical prowess to make assertions about the ground state energy of the fast Hamiltonian for different cases of nonlinearity. They obtain in the sublinear case no dependence of the ground state energy on the slow position variable, independently of other parameters in the circuit. From here they conclude that  setting to zero the parasitic capacitance, no matter other parameters, the quantum mechanical and the classical results completely diverge. Hence they claim that in modelling superconducting devices we should  not rely on Kirchhoff's laws and the reduction in number of variables that follow from them. In other words, outright classical elimination of parasitic elements will lead us to the wrong quantum mechanical results.

We do object to this conclusion.

First of all, in the ``going down" process, there are reasons for caution in the presence of small parameters in dynamical systems, but this  regard for caution need not arise solely from quantum mechanics. Above we have shown  \emph{classical} examples in which there is no sensible slow-fast separation. 
Second, this is not a deficit of the Dirac-Bergmann or the Faddeev-Jackiw methods nor of Kirchhoff's laws.  As we have stressed above, the crucial constraints in which bifurcations are identified are  in fact consistency equations for the reduced dynamical system, independently of the technique used to reach them. Additionally, we have shown examples in which the \emph{classical} slow-fast separation, starting from a regularised model, breaks down precisely at those bifurcation points. Dirac--Bergmann  (or Faddeev--Jackiw, for that matter) actually show their value in identifying the bifurcation directly. To cap it all, the constraints in the fully reduced system are not merely Kirchhoff's laws: they require in an essential manner the concrete dynamics at hand, unlike Kirchhoff's laws, that are independent of the details of the one-port or multi-port elements of the circuit. 

Third, located in the hinge between ``going down" and ``going up", we have not discarded and cannot discard the possibility that some consistent classical reduction and posterior quantization disagrees with quantization and adiabatic reduction for the same system. 
But that is not the situation in examples which include the implicit assumptions of RD. Observe that to be able to identify a collision, the classical adiabatic expansion has to be examined on a par with the Born-Oppenheimer method. On doing so for the more relevant example, we show that there is no contradiction in the nonsingular regime.

Fourth, again in the inflection point, the possibility of compact variables was not examined in RD. Both the classical and the quantum models of reduction are crucially changed in this case, and neither are included in RD's analysis, in spite of the sweeping nature of their final conclusion. Even more, we present detailed proof that determines that the analysis of RD is inapplicable to the case of compact variables, and hence also their conclusions, even if they were valid (and they are not), are inapplicable to this case.

Fifth, now in the ``going up" direction, the validity of the classical slow-fast separation must be examined in each case, and so must the adequacy of the Born-Oppenheimer be. The actual computation in RD pertains only to the spectrum of the low energy states, but there is no analysis of  nonadiabatic transitions. In particular, whether the BO assumption is valid beyond the critical point. For clarity, we have not examined this issue in any matter of detail. Yet, given the generality claimed in the conclusions of RD, this has to be an essential part of their proof, if determined to be valid in some form.

Sixth, in setting up the quantum mechanical part of the argument Kirhhoff's laws were implicitly used throughout. As the conclusion is that they are suspect, the implication is that we are in a state of complete lack of tools for the description of superconducting devices in a lumped element fashion.
Yet the phenomenological and experimental state of affairs directly counters this conclusion in RD: those elements that are to be discarded have played a central r\^ole in the successful modelling of several existing devices.

\subsection{Itemized summary}
\begin{figure}[!ht]
	\includegraphics[width=1
\linewidth]{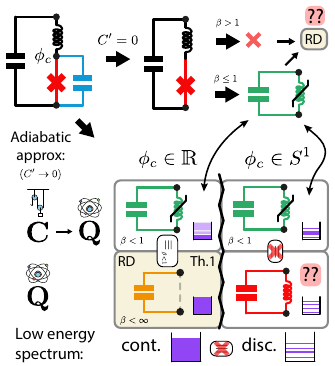}
	\caption{Summary of the results from RD and this work. RD claims the failure of the classical circuit reduction methods (DB, FJ, etc.) due to the impossibility of obtaining standard Hamiltonian dynamics for an ill-posed classical circuit. E.g., for the circuit with a Josephson junction without its parasitic capacitance ($C'=0$), when $\beta>1$. Puzzled by this fact, they explore an alternative route, via a quantum adiabatic approximation. In particular, applying Theorem 1 to the case at hand, (JJ as a nonlinear inductor), they obtain low-energy open circuit dynamics (with an associated continuous spectrum), depicted in orange in the figure, which work, as we have proven, only under the assumption that $\phi_c\in \mathbbm{R}$. In addition, we have proven how their obtained spectrum matches that of the quantization of a model obtained under a classical adiabatic approximation  for $\beta<1$ (left green circuit). In addition, we have shown how those spectra are fundamentally different from those obtained (via adiabatic analysis) under the assumption of $\phi_c\in S^1$, both classically and (naïvely) quantum mechanically (lower right-hand green and red circuits).} 
	\label{fig:Summary_criticism}
\end{figure}

We will now itemize the main points of our previous analysis, as portrayed in Fig.~\ref{fig:Summary_criticism}. They are listed in the order of RD.

\begin{itemize}
\item{\bf RD Sec I:} 
\begin{enumerate}
    \item[-] (minor) Dirac--Bergmann constraint, in this context, is not identical to Kirchhoff's current laws; rather, the dynamical content enters crucially.
\item[-]  (minor) Dirac--Bergmann constraints as incompatible with a regularized approach. As shown above, this is not the case at all. Rather, in the important examples the critical point for Dirac-Bergmann singularity matches the one obtained from regularization.
    \item[-]  The Josephson junction is classified as sublinear. This incorporates the implicit assumption that the range of its flux variable is the full real line.
    \item[-] ``Can be accurately dealt with using the Born-Oppenheimer approximation." There is no analysis as to the validity of the Born-Oppenheimer approximation in the text.
    \item[-] The existence of fixed points is ascribed to the convergence of a perturbation series  for the ground state of the fast Hamiltonian. However, if indeed these fixed points are present, the information on that convergence is not enough to establish their existence.
    
\end{enumerate}
\item {\bf In RD Sec. II:} RD misplaces the failure of the variable reduction of the singular circuit in Fig.~\ref{fig:RD_Circuit}(c) (Fig. 3 in RD) to the Dirac-Bergmann algorithm (and equivalently, to the Faddeev-Jackiw algorithm). The  circuit under scrutiny is ill-defined for $\beta>1$ in the sense of Chua's classification~\cite{Chua:1980}.
    \item {\bf RD Sec. III:} 
    \begin{enumerate}
        \item[-] RD analyzes a set of nonlinear inductors assuming very specific spectra for the quantum operators of Hamiltonian (\ref{eq:Hr_RD_circuit}) (in the alternative assuming very specific topology for the classical configuration space). We prove that their Theorems do not apply when the flux variable $\phi_c$ is considered compact.
        \item[-] The JJ is explicitly classified as type 1(b), instead of 1(a). This shows that their classification does not pertain directly to the element, but rather a to a system after having carried out a Kirchhoff reduction, maligned as the Kirchhoff laws later are. Namely, moving from 1(a) to 1(b) comes about if there is a breaking of time reversal symmetry, due to a flux, that can only enter the problem in presence of a closed loop, i.e. of a circuit beyond the element on its own.
        \item[-] We prove that RD's  resulting low-energy spectrum under the application of Theorem 1 for a Josephson junction (or Th. 2 for other bounded non-symmetric potentials) is not informative, if one considers all operators with real spectrum. In contrast, the nonsingular circuit modeling in previous Sec. II for $\beta<1$ has been succesfully used to explain several experimental data, including \cite{Willsch:2024}. 
        \item[+] On a positive note, their highly technical and valuable Theorems will provide insights in circuits without superconducting islands (Th. 4), e.g., in \emph{fluxonium}-type circuits; as well as in other quantum mechanical systems unequivocally described by extended variables. Further work in this direction will be required to: (a) check the validity of the BO approximation, and (b) to perform a full comparison with a classical adiabatic analysis.
    \end{enumerate}
    \item {\bf RD  Sec. IV:} RD analyzes the SNAIL~\cite{Frattini:2017} circuit under various limits using compact variables or the Bloch theorem (given that there is a trivial mapping here). We point out the  confusion that arises by contrasting  Theorems 1 and 2 (which are only valid for extended variables) and the analysis performed in that section with compact variables.

    \item {\bf RD Sec. V:} RD challenges a universally-used method to reduce variables in describing superconducting circuits, namely Kirchhoff's laws, without providing substantial evidence. The authors overinterpret the results of their theorems and go so far as to dispute, without proof, extremely standard reductions such as the (\emph{linear}) Y-$\Delta$ transformation.
\end{itemize}

\subsection{Final conclusions}

The question of whether constraints in classical mechanics can or not be understood in terms of a limiting process has been present since the inception of analytical mechanics. See for example ch. 4 in \cite{Arnold:2013} where a holonomic constraint is reconstructed as the limit of an infinite harmonic well. In spite of its venerable roots, there has been a recurrent need to reexamine under which conditions this limiting process is indeed equivalent, and more sophisticated asymptotic analyses have been put forward. Without any pretence to being complete, see for instance \cite{Rubin:1957,Froese:2001}. This holds even more true when quantum mechanics comes into play \cite{Froese:2001,Teufel:2003,Wachsmuth:2010}. In the context of superconducting circuits, experimental and theoretical inputs are making it necessary to be more cognizant of the subtleties involving their study, and in particular as to the usefulness and applicability of classical reduction and effective description with quasi-lumped elements. However, we have to be mindful of the fact that classical theorems, such as those in \cite{Rubin:1957}, are formulated with hypotheses that will make them frequently inapplicable to the circuital context.

In this context, the contribution in RD serves greatly to make practitioners aware of the difficulties involved and to increase the necessarily critical outlook on the justification of simple effective models. Additionally, the technical proof of the Theorems in RD will undoubtedly be useful in similar contexts. However, the conclusions in RD, other than a generic raising of awareness to subtleties, are unsupported by the computations and analyses, on the one hand, and, on the other and even more importantly, are misaligned with the current phenomenological and experimental state of play.

On a positive note, Ref.~\cite{Rymarz:2023} has put the focus on a number of open questions in the field of superconducting circuits and also in the wider context of reduction of parasitic elements. Of particular importance we signal that:
\begin{itemize}
    \item 
The extended vs compact discussion is not ended, and very different results/interpretations can be obtained under different assumptions even in very simple circuits.
\item The only clear equivalence (and consensus) between extended/compact perspectives for superconducting circuits, up to now, appears to be when dealing with bounded periodic potentials uncoupled to extended variables. Further inquiry in the lines of \cite{Koch:2009,Murani:2020,ThanhLe:2020,Hakonen:2021,Sonin:2022,Giacomelli:2024} will be of great interest to settle the issue.
\item The existence of bifurcations in parameter space under the process of reduction for the classical case suggests that a similar phenomenon might be present quantum mechanically. The investigation of this question will require more careful adiabatic analysis, in particular in the search of thresholds for bound states in the low energy effective Hamiltonians.
\item If anything, we draw the conclusion from this analysis that careful classical modelling of superconducting circuits is an invaluable tool, even if not infallible, and that the identification of its points of failure will enrich the discussion.
\end{itemize}

\begin{acknowledgments}
We thank Max Hays for identifying a typo in a previous version of the manuscript. I. L. E. acknowledges support by the Basque Government through Grant No. IT1470-22, and project PCI2022-132984 financed by MICIN/AEI/10.13039/501100011033 and the European Union NextGenerationEU/PRTR. A. P.-R. is supported by Juan de la Cierva fellowship FJC2021-047227-I.
\end{acknowledgments}

\appendix

\bibliographystyle{apsrev4-2}
\bibliography{bibliography}

\end{document}